\documentclass{sigchi-ext}
\usepackage[T1]{fontenc}
\usepackage{textcomp}
\usepackage[scaled=.92]{helvet} 
\usepackage{graphicx} 
\usepackage{balance}  
\usepackage{booktabs} 
\usepackage{ccicons}  
\usepackage{ragged2e} 

\def\plainkeywords{Algorithmic Decision-making; deliberation; publics}

\title{``Public(s)-in-the-Loop'': Facilitating Deliberation of Algorithmic Decisions in Contentious Public Policy Domains}

\numberofauthors{4}
\author{
  \alignauthor{
    \textbf{Hong Shen}\\
    \affaddr{Carnegie Mellon University} \\
    \affaddr{Pittsburgh, PA, 15213 USA} \\
    \email{hongs@andrew.cmu.edu} \\
    }
    
    \vfil  \alignauthor{
    \textbf{Ángel Alexander Cabrera}\\
    \affaddr{Carnegie Mellon University} \\
    \affaddr{Pittsburgh, PA, 15213 USA} \\
    \email{cabrera@cmu.edu} \\
    }
    
    \vfil \alignauthor{
    \textbf{Adam Perer}\\
    \affaddr{Carnegie Mellon University} \\
    \affaddr{Pittsburgh, PA, 15213 USA} \\
    \email{adamperer@cmu.edu} \\
    }
     \vfil \alignauthor{
    \textbf{Jason Hong}\\
    \affaddr{Carnegie Mellon University} \\
    \affaddr{Pittsburgh, PA, 15213 USA} \\
    \email{jasonh@cs.cmu.edu} \\
    }
}

\definecolor{linkColor}{RGB}{6,125,233}

\begin{document}

\CopyrightYear{2020}
\setcopyright{rightsretained}
\conferenceinfo{CHI'20,}{April  25--30, 2020, Honolulu, HI, USA}
\isbn{978-1-4503-6819-3/20/04}
\doi{https://doi.org/10.1145/3334480.XXXXXXX}
\copyrightinfo{\acmcopyright}

\maketitle

\newcommand{\jason}[1]{\out{{\small\textcolor{orange}{\bf [*** Jason: #1]}}}}

\newcommand{\alex}[1]{\out{{\small\textcolor{blue}{\bf [*** Alex: #1]}}}}

\RaggedRight{} 

\begin{abstract}
This position paper offers a framework to think about how to better involve human influence in algorithmic decision-making of contentious public policy issues. Drawing from insights in communication literature, we introduce a ``public(s)-in-the-loop'' approach and enumerates three features that are central to this approach: publics as plural political entities, collective decision-making through deliberation, and the construction of publics. It explores how these features might advance our understanding of stakeholder participation in AI design in contentious public policy domains such as recidivism prediction. Finally, it sketches out part of a research agenda for the HCI community to support this work. 
\end{abstract}


\section{Introduction}
With the increasing deployment of algorithmic decision-making systems in many high-stakes sectors in our society, it has become urgent to consider how to better imbue human values into the design of these systems. Recently, HCI scholars have made important contributions towards this direction, for example, by taking a participatory design approach \cite{lee2019webuildai}  or by proposing the method of ``value-sensitive algorithm design'' \cite{zhu2018value}. 

This position paper adds to the growing literature a different and complementary angle by advocating a \textbf{``public(s)-in-the-loop'' approach, i.e., by engaging and facilitating wider public participation in the deliberation of algorithmic decisions}. It argues that this approach is particularly useful in thinking about how to better involve human influence in algorithmic decisions toward highly contentious public policy issues, when large groups of people with diverse perspectives and competing interests are impacted and when there is pervasive disagreement but no universally applicable standard to settle such disagreement. Drawing from communication literature, especially the literature on public sphere, it helps expand the existing conceptual toolkit by adding three important features: publics as plural political entities, collective decision-making through deliberation, and the construction of publics. 



\section{A ``Public(s)-in-the-Loop'' Approach}
In this section, we enumerate three features a ``public(s)-in-the-loop'' approach introduces to our conceptual toolkit. 

\subsection {Publics as plural political entities}
When Habermas \cite{habermas1991structural} first developed the influential concept of the ``public sphere,'' it refered to a historical bourgeois social space that emerged in 18th century Europe where private citizens came together to discuss and debate public issues. Later on, this concept was critiqued for its exclusion of other members of the public, such as women and workers, and various counterpublics have been proposed \cite{fraser1990rethinking}. 

It is important, therefore, to take a pluralistic stance in conceptualizing the social category of ``public(s)''. Instead of a single unified public, scholars have argued that there are multiple different and competing publics \cite{dewey1927public, fraser1990rethinking}. Such a pluralistic stance, on the one hand, suggests a social category that is broad and inclusive. On the other hand, it also indicates the inherent differences, competing interests, and power dynamics among various social groups. 

\subsection{Collective decision-making through deliberation}
Humans are inherently social animals and they often make decisions collectively. In many existing works, human values in AI systems are understood as \textit{individual moral dilemma} and are calculated through aggregations of \textit{individual preferences} (e.g., ask participants to vote whether a self-driving car should kill a baby or a grandma).

Conceptualizing those humans as publics, however, offers an alternative perspective. Scholars of the public sphere \cite{habermas1991structural} have long argued the importance of communication in collective decision-making. One such communicative practice in a liberal democracy is deliberation. Deliberation refers to an approach to politics in which lay people, not just experts, are involved in political decision-making through the exchange of ideas and perspectives via rational discourse \cite{cavalier2011approaching}. Through deliberation, different members of publics will have the opportunities to understand each other's perspectives, challenge one another to think in new ways, and learn from those who are most adversely affected. 

It is important to note that consensus might not be the end goal of deliberation. Mouffe's theories of agnostic pluralism \cite{mouffe1999deliberative} remind us of the importance of radical differences in the practice of democracy. Instead of prioritizing consensus, therefore, we need to broaden our definition of communication practice here to include contentious expression. 

\subsection{The ``construction'' of publics}
Finally, the concept of publics also indicates that there is a formation process. In particular, publics are conceptualized not as pre-existed or fixed social groups but are strangers brought together -- or ``constructed'' -- through and around issues of public interest \cite{dewey1927public}.

Scholars (e.g.,\cite{benkler2006wealth, boyd2010social}) have discussed how digital technologies have enabled both new opportunities and created new problems for constructing ``networked publics'' or ``networked public sphere''. Previous forms of publics have suffered from constraints like physical space, communication speed, archiving and searching. A ``networked public sphere,'' therefore, might have advantages in reaching an even wider public through accessibility; meanwhile, it might also give rise to new problems, like bots or disinformation.  



\section{Using the framework for analysis}
To illustrate how the above three features might advance our understanding of stakeholder participation in AI design in contentious public policy domains, think of the debate on which fairness measures are most appropriate for the recidivism prediction algorithm COMPAS \cite{narayanan_arvind_21_2018}. 

Applying the first feature to the case, the concept of publics highlights the competing political interests among multiple social groups in choosing the ``appropriate'' fairness measure. It thus will not try to find out the ``right'' measure or calculate the majority vote but rather to recognize and expose various competing interests and conflicts first (e.g., decision makers might care more about accuracy while defendants might care more about the false positive rate \cite{narayanan_arvind_21_2018}). 

The second feature of collective decision-making adds to the discussion the importance of creating a communication space to support public deliberation and debate on such algorithmic systems. A consensus may or may not be reached at the end, but through public deliberation, members of publics will be able to learn about each other's perspectives (e.g., why do you care more about the false positive rate?) and a more acceptable solution might emerge. 


Finally, the third feature of ``publics as constructed'' reminds us the importance of bringing members of different publics together around issues of shared interests. We have the opportunity to create critical intervention in this space by exposing the often invisible tensions, conflicts and politics encoded in these seemingly neutral algorithms and raise better public awareness. 

\section{An HCI Research Agenda}
Here, we sketch out part of a research agenda for the HCI community to support this work. 

\subsection{Develop non-expert-oriented toolkits for Explainable AI}
Past work in Explainable AI has primarily focused on how to better support \textit{expert} understanding of ML model \cite{biran2017explanation}, including technical experts (e.g., data scientists) and domain experts (e.g., doctors). Our framework highlights the importance of developing non-expert-oriented toolkits to enable layman's understanding and evaluation of AI systems. Different from ``experts'' and ``domain experts,'' members of publics lack technical training and domain knowledge and have very little time and resources. This presents a distinctive design requirement. For example, can we develop more intuitive and usable interfaces to help them understand the trade-offs of different fairness metrics, comprehend the real world impacts of a ML model, and support their subjective and social evaluation of an AI system? Previous lessons from usable privacy and security might offer help in this regard.

\subsection{Construct communication space for collective decision-making}
Past research in HCI has explored how to better engage citizens in  policy-making \cite{mahyar2018communitycrit}. Our framework highlights the importance of further extending this line of work into algorithmic decisions. Instead of aggregating individual preference, we need develop tools and systems to support deliberation and enable collective decision-making. For example, instead of asking participants to vote, we can ask them to collectively write a policy proposal to demonstrate their understanding and appreciation of each other's perspective. We can also design measures and conduct pre and post tests to evaluate if the deliberation process have influenced people's decisions. 


\subsection{Create interventions for constructing algorithmic publics}
Design scholars have argued that the products and processes of design might contribute to the construction of publics by making invisible societal issues visible \cite{disalvo2009design}. Our framework foregrounds such opportunities for bringing people together around algorithmic decisions. This is also something electronic tools might be able to help with. For example, if a system knows demographics of individuals, it could see if outcomes are balanced or representative of society as a whole. A system might deliberately put people from highly diverse backgrounds in online forums (versus a single massive forum).

\section{Conclusion}
In sum, we propose a ``public(s)-in-the-loop'' approach to conceptualize stakeholder participation for AI design in contentious public policy domains. Our framework adds to the existing conceptual toolkit by highlighting the importance of pluralism, deliberation and public formation. 

\balance{} 

\bibliographystyle{SIGCHI-Reference-Format}
\bibliography{publics}

\end{document}